\documentclass[aps,showpacs,amssymb,pra,superscriptaddress,twocolumn,floatfix]{revtex4-1}

\usepackage{graphicx}
\usepackage{amsmath}
\usepackage{hyperref}
\usepackage{dcolumn}
\usepackage{bm}
\usepackage{color}
\usepackage[utf8]{inputenc}
\usepackage[capitalize]{cleveref}

\allowdisplaybreaks[1]

\begin{document}

\title{Duality of bounded and scattering wave systems with local symmetries}

\author{I. Kiorpelidis}
\affiliation{Department of Physics, University of Athens, GR-15771 Athens, Greece}

\author{F.K. Diakonos}
\affiliation{Department of Physics, University of Athens, GR-15771 Athens, Greece}
	
\author{G. Theocharis}
\affiliation{LUNAM Universit\'e, Universit\'e du Maine, CNRS, LAUM UMR 6613, Avenue O.Messiaen, 72085 Le Mans, France}

\author{V. Pagneux}
\affiliation{LUNAM Universit\'e, Universit\'e du Maine, CNRS, LAUM UMR 6613, Avenue O.Messiaen, 72085 Le Mans, France}

\author{O. Richoux}
\affiliation{LUNAM Universit\'e, Universit\'e du Maine, CNRS, LAUM UMR 6613, Avenue O.Messiaen, 72085 Le Mans, France}

\author{P. Schmelcher}
\affiliation{Zentrum f\"ur Optische Quantentechnologien, Universit\"{a}t Hamburg, Luruper Chaussee 149, 22761 Hamburg, Germany}
\affiliation{The Hamburg Centre for Ultrafast Imaging, Universit\"at Hamburg, Luruper Chaussee 149, 22761 Hamburg, Germany}

\author{P.A. Kalozoumis}

\affiliation{Department of Physics, University of Athens, GR-15771 Athens, Greece}
\affiliation{LUNAM Universit\'e, Universit\'e du Maine, CNRS, LAUM UMR 6613, Avenue O.Messiaen, 72085 Le Mans, France}

\date{\today}

\begin{abstract}
	
We investigate the spectral properties of a class of hard-wall bounded systems, described by potentials exhibiting domain-wise different local symmetries. Tuning the distance of the domains with locally symmetric potential from the hard wall boundaries leads to extrema of the eigenenergies. The underlying wavefunction becomes then an eigenstate of the local symmetry transform in each of the domains of local symmetry. These extrema accumulate towards eigenenergies  which do not depend on the position of the potentials inside the walls. They correspond to perfect transmission resonances of the associated scattering setup, obtained by removing the hard walls. We argue that this property characterizes the duality between scattering and bounded systems in the presence of local symmetries. Our findings are illustrated at hand of a numerical example with a potential consisting of two domains of  local symmetry, each one comprised of Dirac $\delta$ barriers.

\end{abstract}

\maketitle

\section{Introduction}\label{I}

The existence of symmetries provides numerous advantages to the study of a physical system, thereby yielding significant fundamental and phenomenological insights.
The usual practice for most of the studied systems is to assume a global symmetry, i.e. the symmetry holds for the complete system under consideration. In this case, important properties, such as the band structure in periodic settings~\cite{Bloch} or the classification into even and odd eigenstates for systems with global reflection (inversion) symmetry~\cite{Zettili} can be extracted. However, due to the finite size of a realistic system as well as due to the inevitable existence of defects, a globally valid symmetry constitutes an idealized scenario in nature. On the other hand, exact or approximate symmetries which exist in restricted spatial subdomains of a larger system are frequently met. Such spatially localized symmetries can be intrinsic in complex physical systems such as quasicrystals~\cite{Shechtman,Widom,Verberck}, partially disordered matter~\cite{Wochner}, large molecules~\cite{Grzeskowiak,Pascal} or in biological materials~\cite{Gipson}.    

Furthermore, contemporary technology requires structures with specialized properties which are not always possible to achieve in the presence of generic characteristics such as a global symmetry or total disorder. In such cases, local symmetries can be present by design, providing tailored properties and enhanced control in photonic multilayers~\cite{Macia,Zhukovsky,Peng}, semiconductor superlattices~\cite{Ferry}, magnonic systems~\cite{Hsueh}, as well as acoustic~\cite{Hao,Aynaou,King} and phononic~\cite{Hladky,Tamura,Mishra} structures.With the term \textit{local symmetries} (LS) we refer to symmetries which are valid in spatial subdomains of the complete (embedding) space and one possible way is to consider them as \textit{remnants} of a broken global symmetry.

The foundations of local symmetries and their impact in a variety of scattering systems have been investigated in a sequence of recent works. A rigorous mathematical framework for the description of symmetry breaking leading to local symmetries has been developed in~\cite{Kalozoumis2014,Kalozoumis2015b,Zambetakis2016}, where nonlocal invariant currents have been identified as remnants of broken global symmetries. In~\cite{Morfonios2014} it was shown that the long-range order and complexity of lattice potentials generated by well-known binary aperiodic one-dimensional sequences can be encapsulated within their local symmetry structure, while in~\cite{Wulf2016} the case of driven lattices was discussed.  
The scattering properties of quantum and photonic aperiodic structures were discussed in~\cite{Kalozoumis2013a,Kalozoumis2013b}, answering the puzzling question about the existence of perfect transmission resonances in aperiodic systems and also providing a classification scheme with respect to their kind. These theoretical findings were firstly experimentally verified in~\cite{Kalozoumis2015} in the framework of acoustic waveguides. Apart from continuous scattering systems, the impact of local symmetries has been also investigated in the framework of discrete systems~\cite{Morfonios2017}. In this context, the local symmetry partitioning revealed new possibilities for the design of flat bands and compact localized states~\cite{Rontgen2018a}. Very recently, it was shown that the existence of local symmetries plays a crucial role in the real space control of edge states in aperiodic chains~\cite{Rontgen2018b} as well as in the wave delocalization and transport between disorder and quasi-periodicity~\cite{Morfonios2018}. Thus, the concept of local symmetries, even being a recent one, has already led to a rich phenomenology and revealed properties of fundamental importance. 

Even though local symmetries in continuous scattering and discrete systems have been extensively investigated in the aforementioned works, their consequences and effects in continuous bounded systems remains unexplored. The link between bounded and unbounded systems is a long standing subject of study in quantum mechanics and in the more general context of wave physics ~\cite{Ambichl2013}. Several methods have been employed to describe how a bounded system is connected to its open counterpart~\cite{Stockmann2002}, since spectral properties usually can be measured when the system is coupled to an environment.
Nonetheless, this link has never been studied under the prism of local symmetries.

As a first step in this direction we explore in this work the properties of one-dimensional bounded systems with two locally symmetric  potential barriers, focusing on the case of local reflection symmetries. We define as \textit{setup} the two locally symmetric potential barriers along with the distance which separates them, while the term \textit{system} is used to describe the entire potential landscape consisting of both the setup and the bounding hard walls.
Tuning the distance of the setup from the left hard wall, we prove the existence of spectral extrema where the mirror symmetry of the wavefunction is restored inside each reflection symmetric potential barrier. We also establish a link between the spectral properties of a generic bounded system with two domains of local symmetry  and the properties of the respective scattering system. In particular, we find that certain eigenenergies of the bounded system correspond to the energies where perfect transmission resonances (PTRs) manifest in the transmittance of its scattering counterpart and we prove that these eigenenergies are unaffected by the position change of the setup inside the walls. These theoretical findings are numerically verified for a system with two domains of local reflection symmetry comprised of Dirac $\delta$-barriers.

The paper is organized as follows: In Section~\ref{II} we summarize the key ingredients and present some basic results of scattering theory in systems with local symmetries. Also we introduce the setups which we will employ, both in the scattering and in the bounded context. In Section~\ref{IV} we focus on the properties of a bounded system with local symmetries comprised of Dirac $\delta$ barriers and discuss the relevant properties. We also discuss the connection between certain bounded states and PTRs. In Section~\ref{V} we generalize rigorously our results for a generic system with two domains of local symmetry of arbitrary potential shape.  Our results are summarized in Section~\ref{VI}.

\section{Scattering in systems with local symmetries - An overview}\label{II}

\subsection{Perfect transmission resonances}

Scattering potentials possessing a global mirror symmetry and the possibility of the occurrence of perfect transmission resonances (PTRs) have been directly linked to each other in several studies~\cite{Huang1999,Huang2001}. On the other hand, the lack of such a symmetry usually leads to a nonvanishing reflection of an incoming scattering wave. However, the existence of PTRs in aperiodic~\cite{Peng} structures possessing no global mirror symmetry has been reported.
Recently, we established~\cite{Kalozoumis2013b} a classification of the possible PTRs which occur in non-globally symmetric systems. In particular, for a system with local symmetries the PTRs can be classified according to the symmetry of the wavefunction modulus $u(x)=|\psi(x)|^2$. If $u(x)$ is reflection symmetric within the domains of local reflection symmetry then it is called a \textit{symmetric} PTR ($s$-PTR) whereas if $u(x)$ does not obey this local symmetry is called \textit{asymmetric} PTR ($a$-PTR). In the $s$-PTR case each domain of local symmetry is individually transparent. For $a$-PTRs the system is transparent only as a whole.

\begin{figure}[h!]
\begin{center}
\includegraphics[width=0.9\columnwidth]{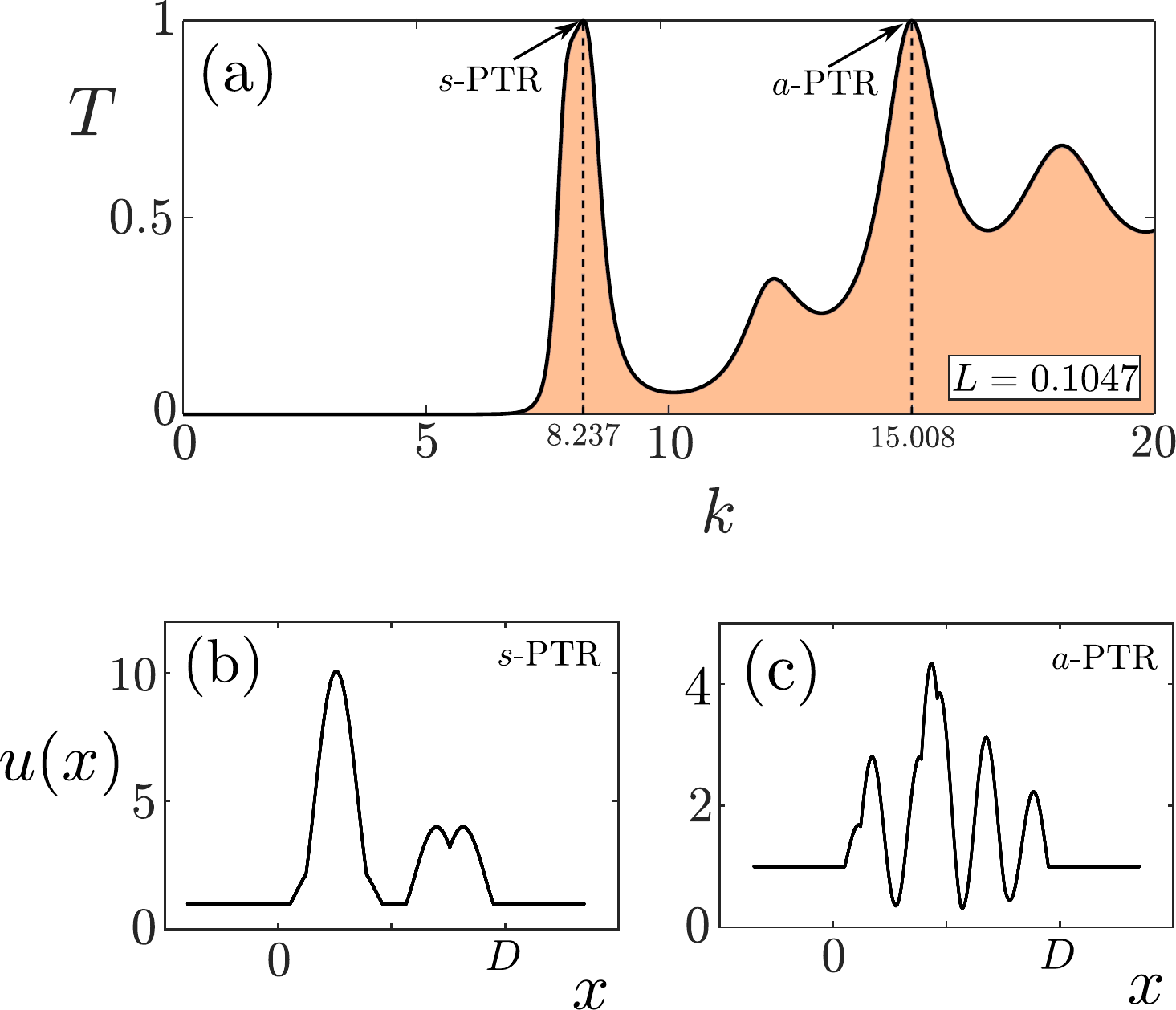}
\end{center}
\caption{(a) Transmittance of the setup shown in Fig.~\ref{fig1} (a). The two peaks correspond to a $s$- and an $a$-PTR, respectively. (b) and (c) illustrate the magnitude to the wavefunction at the wavenumbers of those PTRs. Here $D=1.0016$ (and $L=0.1047$). \label{fig1}}
\end{figure}

Figure~\ref{fig1} (a) shows the transmittance of the setup shown in Fig.~\ref{fig2} (a). The two peaks correspond to an $s$- and an $a$-PTR, as their wavefunction moduli indicate in Fig.~\ref{fig1} (b) and (c), respectively. Note here, that the PTRs shown in  Fig.~\ref{fig1} (a) do not occur by chance. The setup is designed according to prescription based on local symmetries and the parameters (indicated in Fig.~\ref{fig2}) are suitably tuned in order to  emerge at the specific frequencies. Also different kind of tuning is required for an $a$- and a $s$-PTR, respectively.  This design technique and a thorough investigation of the scattering properties of the system which corresponds to the transmittance shown in Fig.~\ref{fig1} (a)  can be found in Ref.~\cite{Kalozoumis2015}. 

\subsection{Symmetry induced invariant currents}

Another important finding of our theoretical framework on local symmetries is the existence of symmetry-induced currents which are spatially invariant in domains where a certain symmetry i.e. reflection, translation or $\mathcal{PT}$ symmetry is present. Employing a generic wave mechanical framework we consider a generalized Helmholtz equation $\psi''(x) + U(x) \psi(x)=0$, where $U(x)$ is the generalized potential. In this framework, it is possible to treat in a unified way different wave mechanical systems of e.g. photonic, acoustic and quantum mechanical origin. Assuming that the potential $U(x)$ obeys a symmetry transformation $U(x)=U[F(x)]$ within a domain $\mathcal{D}\subseteq \mathbb{R}$ ($\mathcal{D} = \mathbb{R}$ corresponds to a global symmetry), it can be shown that a spatially invariant, nonlocal current exists,
\begin{equation}
\mathcal{Q}=\frac{1}{2i}[\sigma \psi(x)\psi'(\tilde{x})-\psi(\tilde{x})\psi'(x)] = \text{const} ~~~ \forall x,\tilde{x}\in\mathcal{D}.
\label{Q} 
\end{equation}  
This quantity plays a fundamental twofold role. It provides the tool to systematically describe the breaking of discrete symmetries, while it also generalizes the Bloch and parity theorems  for systems with broken translation and reflection symmetry, respectively~\cite{Kalozoumis2014}. The quantity $\mathcal{Q}$ is of central importance for this study. Note, that in bounded systems -since the wavefunction is real- only the invariant current $\mathcal{Q}$ exists. On the other hand, in scattering systems where the wavefunction $\psi(x)$ is complex an additional invariant quantity
\begin{equation}
\mathcal{\tilde{Q}}=\frac{1}{2i}[\sigma \psi^*(x)\psi'(\tilde{x})-\psi(\tilde{x})\psi'^{*}(x)] = \text{const} ~~~ \forall x,\tilde{x}\in\mathcal{D} 
\end{equation}  
emerges.

\subsection{Description of the setup}

\begin{figure}[h!]
\begin{center}
\includegraphics[width=0.9\columnwidth]{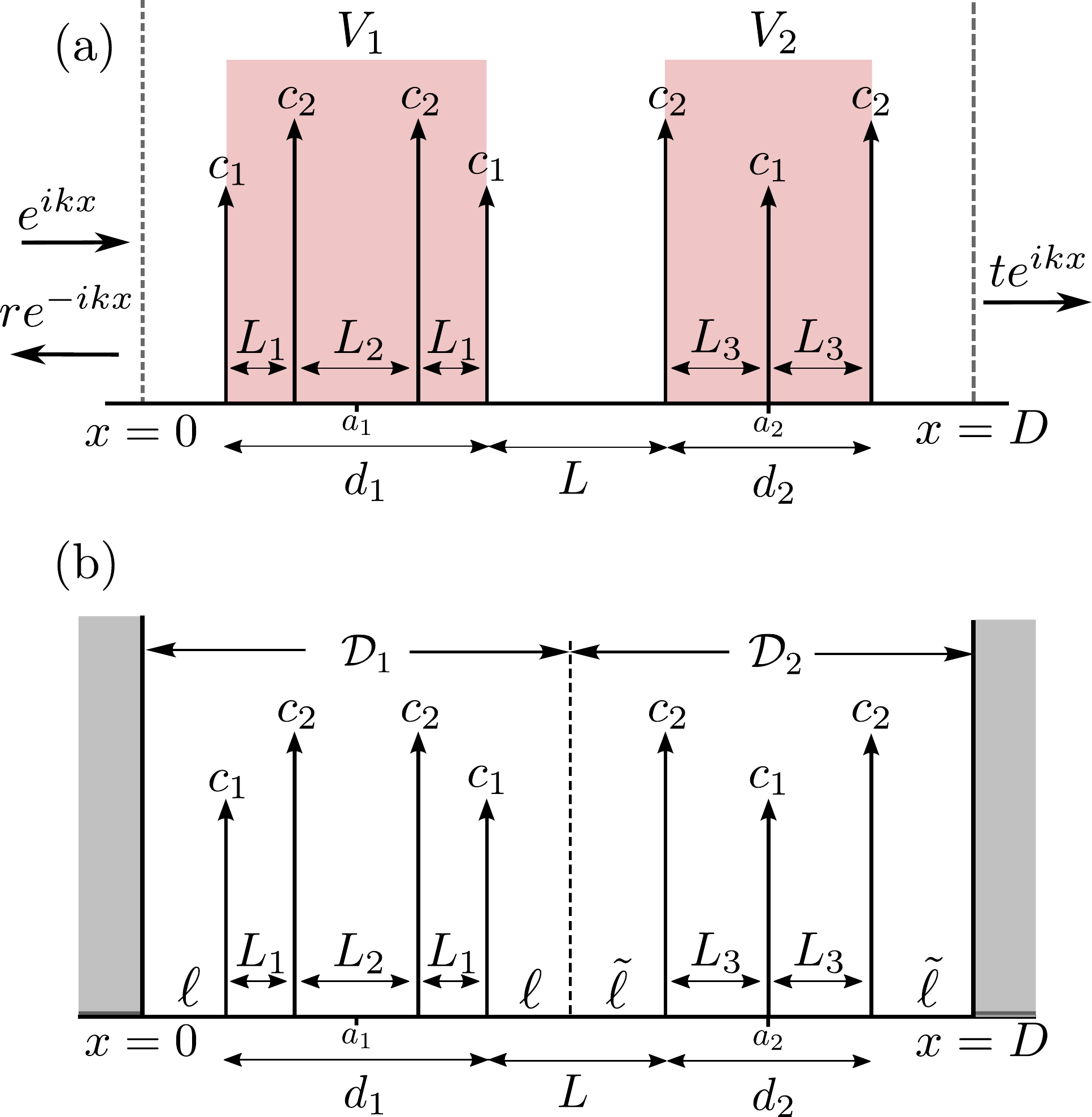}
\end{center}
\caption{\label{fig2} (a) Scattering and (b) bounded systems with local symmetries. Each potential subparts $V_{1},~V_{2}$ is reflection symmetric. The distances between the delta functions are equal to $L_1=0.07$, $L_2=0.267099$, $L_3=0.192539$ and the strengths of the $\delta$ barriers are equal to $c_1=7.8886$ and $c_2=12.3414$.}
\end{figure}

Let us now describe the setup which we will use throughout this work. Figure~\ref{fig2} (a) illustrates a scattering system comprised of seven Dirac $\delta$ barriers, forming two reflection symmetric potential subparts denoted as $V_{1}$ and $V_{2}$ (coloured areas). The lengths of $V_{1}$ and $V_{2}$ are $d_{1}$ and $d_{2}$, respectively, while their separating distance is $L$. Moreover, $a_{1}$ and $a_{2}$ stand for the positions of the reflection centers of each subpart. The parameters $c_{i}, i=1,2$ represent the strength of the $\delta$ barriers and $r,~t$ are the reflection and transmission coefficients. A detailed study of the scattering properties of this system and their experimental verification can be found in Ref.~\cite{Kalozoumis2015}. On the other hand, Fig.~\ref{fig2} (b) shows the corresponding bounded system, where the aforementioned setup is delimited by hard wall boundaries. The general characteristics remain the same. However, the distance $\ell$ which determines the distance from the left wall plays a crucial role and in the following will serve as our tuning parameter. In order to preserve the local symmetries of the system for any value of $\ell$ -namely the domains $\mathcal{D}_{1}$ and $\mathcal{D}_{2}$ being always reflection symmetric- it should hold $L=\ell+\tilde{\ell}$ [see Fig.~\ref{fig2} (b)].

\section{Local symmetries in bounded systems}\label{IV}

\begin{figure*}
\begin{center}
\includegraphics[width=0.9\textwidth]{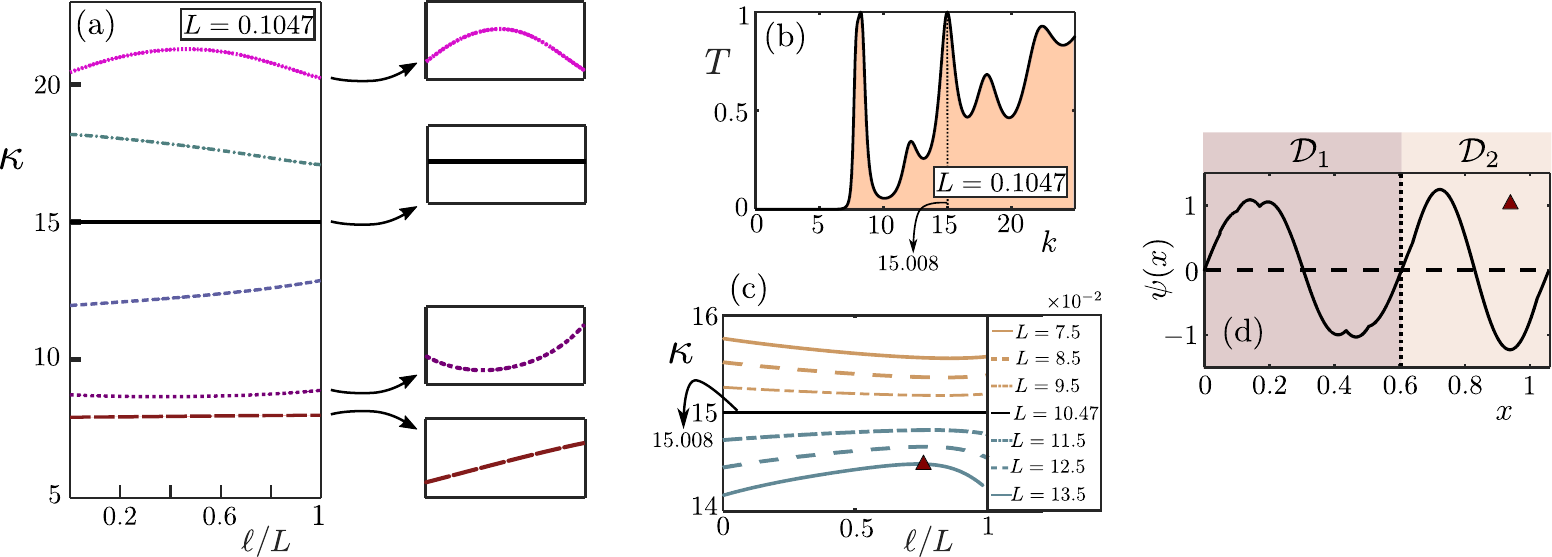}
\end{center}
\caption{(a) Spectrum showing the first 6 states of the bounded system shown in Fig~\ref{fig2} (b) for $L=0.1047$. As the tuning parameter $\ell$ varies in the range $[0,L]$ the wavenumber $\kappa$ varies continuously. The insets zoom into the curves to provide a better resolution. (b) Transmittance for the corresponding open system. The dashed line at $k=15.008$ indicates an $a$-PTR. (c) The fourth state of spectrum of the bounded system for different $L$ values. For $L=0.1047$ a duality between the open and closed systems occurs connecting the PTR wavenumber $k=15.008$ and the invariant bounded state $\kappa=15.008$.  (d) Wave function which follows the local symmetries of the setup at the maximum indicated with the triangle.  \label{fig3}}
\end{figure*}

Several connections  between bounded and scattering systems can be investigated. In this work we focus on bounded systems and how they can be linked to their scattering counterparts from the perspective of \textit{PTRs} and \textit{local symmetries}. 
To this end, we consider the bounded version of our system as shown in Fig.~\ref{fig2} (b). 
We remind the reader that for the lengths $\ell,~\tilde{\ell}$ it holds $L=\ell+\tilde{\ell}$. With this choice -and employing $\ell$ as our tuning parameter with $\ell \in [0,L]$- we ensure that for any value of $\ell$ the system is always decomposable into two locally symmetric domains $\mathcal{D}_{1}$ and $\mathcal{D}_{2}$. Keeping $L$ fixed and varying $\ell$ we expect that the spectrum of the allowed wavenumbers will change continuously. During this variation we will examine the spectral properties which emerge due to the local symmetries.

Figure~\ref{fig3} (a) shows the first six states of the bounded system [see Fig~\ref{fig2} (b)] for $L=0.1047$ and how these change as $\ell/L$ varies within the range $[0,1]$. The small sub-figures on the right provide a zoom into the selected curves for better resolution. A striking characteristic is the perfectly flat, black solid line  which indicates that the wavenumber $\kappa$ remains unaffected by the position of the setup inside the box. This is an important observation because eigenenergies with this property emerge in the respective scattering counterpart as PTRs, offering the ground for establishing a duality between open and closed systems.
 Figure~\ref{fig3} (b) shows the transmittance of the corresponding scattering system. The dashed line indicates an $a$-PTR peak at $k=15.008$. In Fig.~\ref{fig3} (c) we focus on the fourth state of (a) and examine its behaviour for several $L$ values in the vicinity of $L=0.1047$. For  $L<0.1047$ each $\kappa(\ell/L)$ curve exhibits a minimum. Exactly at $L=0.1047$ the wavenumber $\kappa=15.008$ becomes invariant with respect to the position of the setup within the walls and this manifests through the flat solid line. Remarkably, this $\kappa$ value is identical to the $k$ value of the $a$-PTR in the transmittance of (b). For $L>0.1047$  the $\kappa(\ell/L)$ curves exhibit maxima. All extrema approach the flat line of $L=0.1047$.

\begin{figure*}
\begin{center}
\includegraphics[width=0.9\textwidth]{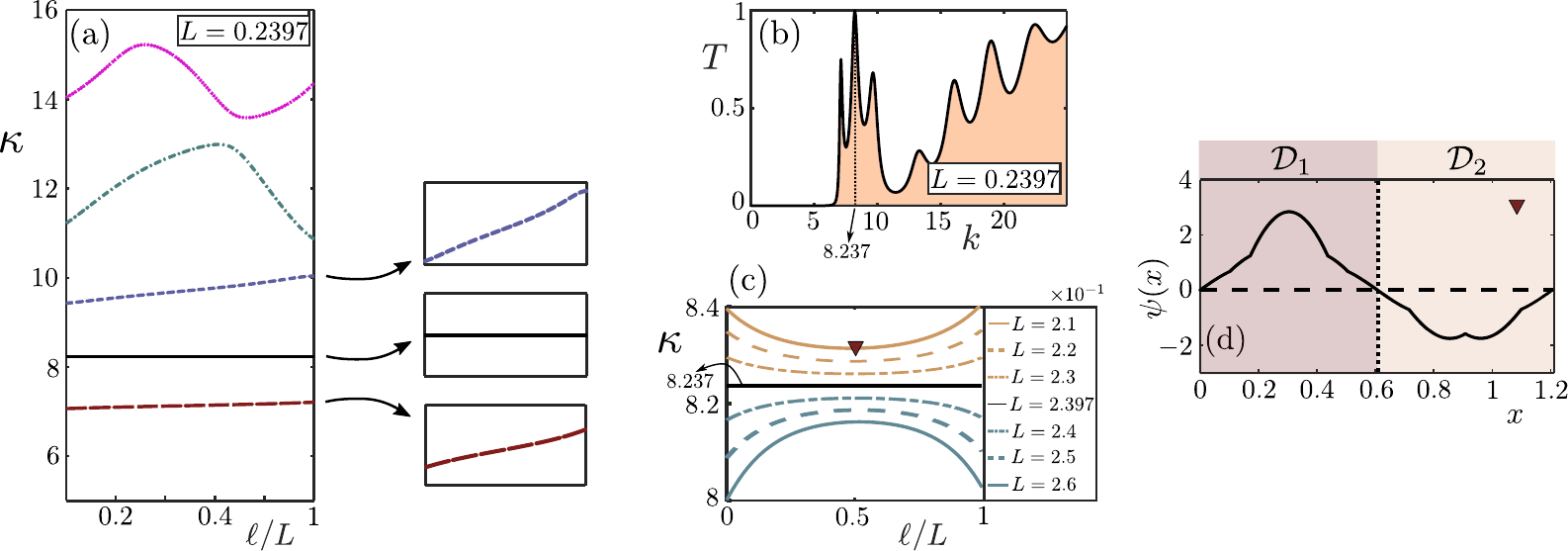}
\end{center}
\caption{(a) Spectrum showing  5 states of the bounded system shown in Fig~\ref{fig2} (b), this time for $L=0.2397$. As  $\ell$ varies in  $[0,L]$ the wavenumber $\kappa$ varies continuously. (b) Transmittance for the corresponding open system. The dashed line at $k=8.237$ indicates an $s$-PTR. (c) The second state of spectrum of the bounded system for different $L$ values. For $L=0.2397$ a duality between the open and closed systems occurs connecting the PTR wavenumber $k=8.237$ and the invariant bounded state $\kappa=8.237$.  (d) Wave function which follows the local symmetries of the setup at the minimum indicated with the down triangle.  \label{fig4}}
\end{figure*}

It is obvious that for each $L$ value, both the spectrum of the bounded system and the transmittance of the scattering system will change. Nevertheless, the aforementioned correspondences can be identified. In Fig.~\ref{fig4}  the respective properties of the system  for  $L=0.239$ are discussed. Figure~\ref{fig4} (a) shows  five states of the spectrum and how these change as $\ell/L$ varies within the range $[0,1]$.  The eigenstate which is invariant under $\ell/L$ shifts corresponds to the $s$-PTR shown in the transmittance of Fig.~\ref{fig4} (b).  In Fig.~\ref{fig4} (c) the dependence of the second state as $L$ changes is shown.  The pattern is the same as in the previous example. For $L=0.239$ the wavenumber $\kappa=8.237$ remains constant as the $\ell/L$ changes. For this $L$ value the transmittance exhibits a $s$-PTR peak at the same wavenumber $k=8.237$. For $L \neq 0.239$, $\kappa(\ell/L)$ possess extrema, which `saturate' to the flat line.

Another very interesting property which is observed here relates the extrema of the $\kappa(\ell/L)$ curves with the local symmetries of the setup and the form of the wavefunction.  In particular, at every extremum of the $\kappa(\ell/L)$ curves,   the wavefunction becomes an eigenstate of the local reflection symmetry transform and follows the local symmetries of the setup. 
Figures~\ref{fig3} (c) and (d) illustrate this case. In particular, Fig.~\ref{fig3} (d) shows the wavefunction at the maximum of the curve for $L=0.0135$ (the correspondence is indicated by the  $\triangle$). It is clear that the wavefunction is (locally) parity definite within the local symmetry domains $\mathcal{D}_{1}$ and  $\mathcal{D}_{2}$ of the setup. The same holds for the system in Fig.~\ref{fig4}.  The wavefunction in Fig.~\ref{fig3} (d) corresponds to the minimum of the   $\kappa(\ell/L)$ curve for $L=0.21$ (see the $\triangledown$)  and is locally parity definite following the symmetries  of the $\mathcal{D}_{1} $ and $\mathcal{D}_{2}$. This correspondence between the $\kappa(\ell/L)$ extrema and the local symmetry properties of the wavefunction provides the -possibly- first systematic attempt to investigate the manifestation of local symmetries in bounded systems. 

\begin{figure}
\begin{center}
\includegraphics[width=0.99\columnwidth]{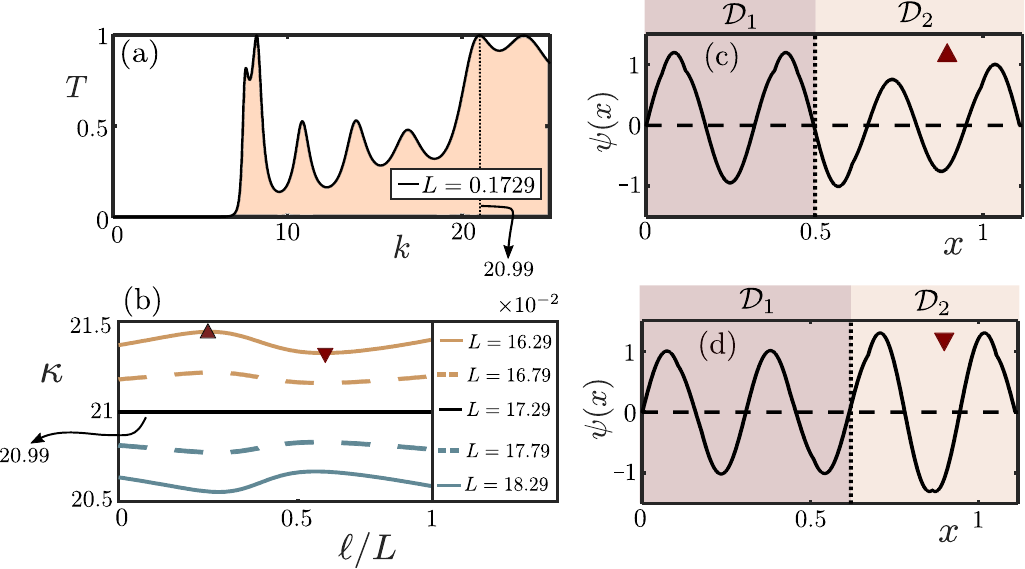}
\caption{(a) Transmittance of the setup shown in Fig.~\ref{fig1} (a) for $L=0.1729$. The dashed line corresponds to a PTR which occurs for the specific $L=0.1729$ choice at $k=20.99$. (b) Variation of the seventh state with $\ell/L$ of the spectrum of the respective bounded system for several $L$ choices around $L=0.1729$. Each $\kappa$ vs $\ell/L$  curve exhibits one minimum and one maximum. For $L=0.1729$ the wavenumber ($\kappa=20.99$) is independent of $\ell$ with a value coinciding with the wavenumber $k$ of the $a$-PTR. (c), (d) Wavefunctions at the maximum and minimun of the first curve (indicated by up and down triangles). Here the wavefunction becomes an eigenstate of the local symmetry transform.   \label{fig5}} 
\end{center}
\end{figure}

In fact, there are $\kappa(\ell/L)$ curves which may possess more than one extrema, as those shown in Fig.~\ref{fig5} (b). These curves correspond to the seventh state of a bounded system for several $L$ values around $L=0.1729$. The flat line occurs for this $L$ at $\kappa=20.9875$ and again benchmarks the appearance of an $a$-PTR in the corresponding scattering setup, as indicated in Fig.~\ref{fig5} (a), a finding which supports the duality between open and closed systems at the PTR wavenumbers. Figures~\ref{fig5} (c), (d) illustrate the wavefunction at the two extrema of the first curve (marked with the up and down triangles). In both cases, it becomes parity definite within the two domains of local symmetry $\mathcal{D}_{1}$ and $\mathcal{D}_{2}$. Note that the fifth state in Fig.~\ref{fig4} (a) also possesses two extrema. However, we showed the case of a different setup in order to stress further the correspondence between  PTRs and translationally invariant bound states.

Note that this bound-scattering duality and the manifestation of local symmetries on the extrema are not system specific. Our conclusions will be rigorously proven and generalized for systems with two locally symmetric potentials domains of arbitrary shape in the following section. 
To conclude this Section we summarize our key findings in the following statement which holds for the general case:
``Consider a bounded system which consists of two domains  of local symmetry $\mathcal{D}_{1},~\mathcal{D}_{2}$, each one with a reflection symmetric potential $V_{1},~V_{2}$  of arbitrary shape and finite support. Between this system and its scattering counterpart  the following duality holds:
 (i) Starting from a bounded system: \textit{If a bound state with wavenumber $\kappa$ is invariant with respect to translations of the setup inside the cavity, then it corresponds to a PTR ($a$ or $s$ ) in the corresponding scattering system with incoming wavenumber $k=\kappa$}. (ii) Starting from a scattering system: \textit{The existence of an  $a$-PTR in a scattering system at $k$ is equivalent to a bound state with wavenumber $\kappa=k$ which is invariant under translations of the (same) setup inside the cavity.} The reason that we discriminate $s$- from $a$-PTRs is because the wavenumber $k$  of an $a$-PTR will always emerge  as an eigenstate with wavenumber $\kappa$ in the corresponding bounded system. This one-to-one correspondence between scattering and bounded systems exists because the $a$-PTR occurs for a specific distance $L$ between the two locally symmetric scatterers. For the $s$-PTR, on the other hand, this one-to-one correspondence between scattering and bounded systems would not be possible because it appears in the transmittance for any distance $L$ between the locally symmetric scatterers~\cite{Kalozoumis2013b}. In this case, all bounded systems' spectra for all different $L$ values  would have an eigenstate at the same $\kappa$, which is not possible. Nevertheless, if the bounded system has an eigenstate at a $\kappa$ value which coincides with the $k$ wavenumber  of an $s$-PTR of its scattering counterpart, then the equivalence between the $s$-PTR and the bound state translation invariance is preserved (case shown in Fig.~\ref{fig4}). (iii)\textit{ All extrema which emerge in the $\kappa(\ell/L)$ curves correspond to states which are eigenstates of the the local reflection symmetry transform} and the wavefunction is parity definite inside $\mathcal{D}_{1}$ and $\mathcal{D}_{2}$.''

\section{Generalization for an arbitrary bounded system with two domains of local symmetry}\label{V}

\begin{figure}
\begin{center}
\includegraphics[width=0.9\columnwidth]{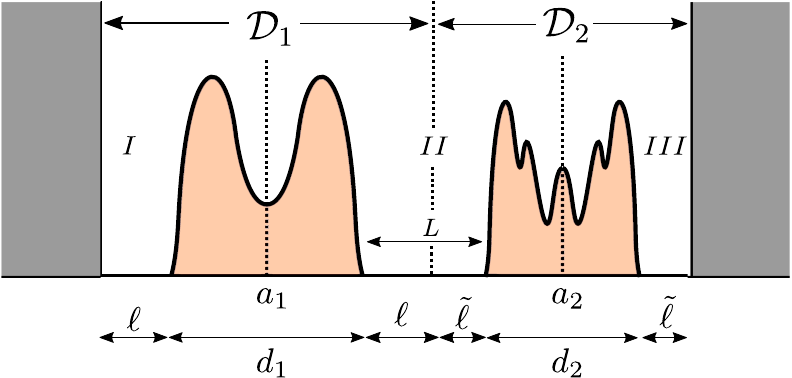}
\caption{\label{fig6}Schematic of a bounded system with hard wall boundary conditions containing two domains of local symmetry $\mathcal{D}_{1}, \mathcal{D}_{2}$. In the regions $I,II,III$ the potential vanishes.}
\end{center}
\end{figure}

In this Section we will generalize the results presented above for arbitrary bounded systems with two domains of local symmetry. 
%
In order to prove the above statement we employ the transfer matrix (TM) approach to connect the wave fields in the regions $I$, $II$ and $III$ of the system, as shown in Fig.~\ref{fig6}.  Since $I$, $II$ and $III$ are potential free regions, the wavefunction will be of the form,
\begin{equation}
\label{plane_waves} \psi_{m}(x)=A_{m} e^{i \kappa x} + B_{m} e^{-i \kappa  x}~~~;~~~ m=I,~II,~III.
\end{equation}
The connection between $\psi_{I}$ and $\psi_{II}$ is provided by the TM, which reads for a Hermitian system,
\begin{equation}
\label{TM} M_{\mathcal{D}_{j}}=
\begin{bmatrix}
w_{j} &  z_{j}\\
z_{j}^{*} & w_{j}^{*}
\end{bmatrix} ~~~;~~~j=1,2.
\end{equation}
Here $j=1,2$ corresponds to the TMs of the LS symmetric subparts defined on $\mathcal{D}_{1}$ and $\mathcal{D}_{2}$, respectively. For the TM elements it holds that $w_{j}=1/t_{j}$ and $z_{j}=r_{j}^{*}/t_{j}^{*}$, where $t_{j}$ and $r_{j}$ are the transmission and reflection amplitudes of the $j$-th potential unit, respectively. 

Positioning the first wall at $x=0$, the wavefunction should be zero there i.e. $\psi(0)=0$, leading to the condition $A_{I}=-B_{I}$ and subsequently to,
\begin{equation}
\label{proof1} \frac{A_{II}}{B_{II}}=-\frac{z_{1c}e^{-2i\kappa  a_{1}}+w_{1c}^{*}}{z_{1c}^{*}e^{2i\kappa  a_{1}}+w_{1c}},
\end{equation}
connecting regions $I$ and $II$. Note here that the index $c$ in the TM elements (referring to ``centered'') corresponds to the TM for $\mathcal{D}_{1}$ centered at $x=0$. The shift to the actual position of $\mathcal{D}_{1}$ in our setup is realized by the phases in Eq.~(\ref{proof1}), with $a_{1}$ being the position of the reflection axis of $\mathcal{D}_{1}$.

In the same manner, taking the wavefunction zero at the second wall $\psi(D)=0$, we find the condition $B_{III}=-A_{III}e^{2ikD}$ which, in turn, leads to the relation,
\begin{equation}
\label{proof2}  \frac{A_{II}}{B_{II}}=\frac{w_{2c}-z_{2c}e^{-2i\kappa  a_{2}}e^{2i\kappa  D}}{z_{2c}^{*}e^{2i\kappa  a_{2}}-w_{2c}^{*}e^{2i\kappa  D}},
\end{equation}
connecting the plane wave coefficients in regions $II$ and $III$.  $a_{1}$ and $a_{2}$ correspond to the mirror symmetry center of the two scatterers (see Fig.~\ref{fig6}).
In turn, Eq.~(\ref{proof1}) and Eq.~(\ref{proof2}) yield,
\begin{equation}
\label{proof3} \mathcal{G}=\frac{z_{1c}e^{-2i\kappa a_{1}}+w_{1c}^{*}}{z_{1c}^{*}e^{2i\kappa  a_{1}}+w_{1c}}+\frac{w_{2c}-z_{2c}e^{-2i\kappa  a_{2}}e^{2i\kappa  D}}{z_{2c}^{*}e^{2i\kappa  a_{2}}-w_{2c}^{*}e^{2i\kappa  D}}=0,
\end{equation}
which involves only the wave number $\kappa $ and the characteristic parameters of the system which are included in the TM elements $w_{jc},~z_{jc} ~(j=1,~2)$. In order to facilitate the mathematical computations we consider only the numerator of Eq.~(\ref{proof3}), which we set as $\mathcal{F}$. Since we are interested in the behaviour of $\kappa $ with respect to $\ell$ it is sufficient to calculate the total derivative of $\mathcal{F}$ with respect to $\ell$ and then calculate the derivative $d\kappa /d\ell$
\begin{equation}
\label{proof4} \frac{d\mathcal{F}}{d \ell} = \frac{\partial \mathcal{F}}{\partial \ell} + \frac{\partial\mathcal{F}}{\partial \kappa }  \frac{d \kappa }{d \ell} =0,
\end{equation}
which leads to
\begin{equation}
\frac{d\kappa }{d\ell}=-\dfrac{\frac{\partial \mathcal{F}}{\partial \ell}}{\frac{\partial \mathcal{F}}{\partial \kappa }}.
\end{equation}
Therefore the behaviour of the wavenumber $\kappa$ with respect to $\ell$ can  be investigated  via the term $\partial \mathcal{F}/\partial \ell$. Note that in order to find $d\mathcal{F}/d\ell$ we have expressed $a_{1,2}$ with respect to $\ell,~d_{1},~d_{2}$ and $L$, namely $a_{1}=\ell+d_{1}/2$ and $a_{2}=\ell+d_{1}+L+d_{2}/2$. Then, the latter is written as
\begin{align}
\label{proof5} \partial \mathcal{F}/\partial \ell= 2i\kappa  \left[ e^{i\kappa \mathcal{P}}\left(w_{1c}^{*}z_{2c}^{*}+z_{1c}^{*}w_{2c}e^{-i\kappa  D}\right) \right. \nonumber \\ \left. +  e^{-i \kappa \mathcal{P}} \left(w_{1c} z_{2c}+z_{1c}w_{2c}^{*}e^{i\kappa  D}\right) \right],
\end{align}
where $\mathcal{P}=2 \ell +d_{1}$.
In the following we will show that this equation  has a very instructive form regarding the emergence of local symmetries.  To this end, we employ the existence of the symmetry induced invariant current $\mathcal{Q}$  for reflection symmetry, as defined in Eq.~(\ref{Q}). Given the plane wave solution in the potential free regions $I$ , $II$ and $III$  [see Eq.~(\ref{plane_waves})] we find that the form for $\mathcal{Q}_{1}$ and  $\mathcal{Q}_{2}$ in the LS domains $\mathcal{D}_{1}$ and $\mathcal{D}_{2}$ is,
\begin{subequations}
\begin{equation}
\label{q1} \mathcal{Q}_{1} = \kappa \left(A_{I} A_{II} e^{2i\kappa  a_1} + B_{I} B_{II} e^{-2i\kappa  a_1} \right),  
\end{equation}

\begin{equation}
\label{q2} \mathcal{Q}_{2} = \kappa \left(A_{II} A_{III} e^{2i\kappa  a_2} + B_{II} B_{III} e^{-2i\kappa  a_2} \right),  
\end{equation}
\end{subequations}
where  $A_{j}, B_{j}$ are the plane wave coefficients in regions $I,~II$ and  $III$ and $\kappa$ is the wavenumber of the specific eigenstate. We stress here that $\mathcal{Q}$ can be calculated taking into account only the potential free regions $I,~II,~III$, since -due to the symmetry- it is independent of the exact potential form (see Ref.~\cite{Kalozoumis2014}). Note also that Eqs.~(\ref{q1}),~(\ref{q2}) has no dependence on $x$, signaling its spatial invariance. Focusing on the setup shown in Fig.~\ref{fig1} we find that $\mathcal{Q}_{1}$ for the domain $\mathcal{D}_{1}$ reads as follows
\begin{equation}
\label{q1n} \mathcal{Q}_{1}= kA_{I} B_{II}\left( \frac{A_{II}}{B_{II}} e^{2i\kappa  a_{1}} +  e^{-2i\kappa  a_{1}} \right),  
\end{equation}
where we have used the condition $B_{I}=-A_{I}$.  The behaviour of $\mathcal{Q}_{1}$  determines the LS properties of the wavefunction in the domain $\mathcal{D}_{1}$. In particular, if $\mathcal{Q}_{1}=0$  the wavefunction inside $\mathcal{D}_{1}$ will be parity definite. Substituting Eq.~(\ref{proof1}) into Eq.~(\ref{q1n}) and setting $\mathcal{Q}_{1}=0$ we find
\begin{equation}
\label{q1zero} z_{1c}=\frac{1}{2} \left( w_{1c} e^{-i\kappa  \mathcal{P}} - w_{1c}^{*} e^{i\kappa  \mathcal{P}} \right),
\end{equation}
where we have used the property $z_{1c}=-z_{1c}^{*}$ which holds for the TM of reflection symmetric potentials. 
Following the same procedure for the domain $\mathcal{D}_{2}$ and for $\mathcal{Q}_{2}=0$ we find,
\begin{equation}
\label{q2zero} z_{2c}=\frac{1}{2} \left( w_{2c} e^{i\kappa  \mathcal{P}} e^{-i\kappa  D} - w_{2c}^{*} e^{-i\kappa  \mathcal{P}} e^{\kappa k D} \right).
\end{equation} 
We stress here that when Eqs.~(\ref{q1zero}) and (\ref{q2zero}) hold, then the wavefunction is parity definite inside $\mathcal{D}_{1}$ and $\mathcal{D}_{2}$, respectively. 

The next step is to substitute Eqs.~(\ref{q1zero}),~(\ref{q2zero}) into Eq.~(\ref{proof5}). This allows to focus on the behaviour of the quantity $\partial \mathcal{F}/ \partial \ell$ when the wavefunction is parity definite in both domains $\mathcal{D}_{1},~\mathcal{D}_{2}$.  After some algebraic manipulations, we find that $\partial \mathcal{F}/ \partial \ell=0$ and, in turn, that
\begin{equation}
\label{k_zero} d \kappa  / d \ell = 0.
\end{equation} 
Therefore, the restoration of LS in the wavefunction (i.e. the field is parity definite inside $\mathcal{D}_{1},~\mathcal{D}_{2}$ and $\mathcal{Q}_{1}=\mathcal{Q}_{2}=0$) in any structure comprised of two barriers of arbitrary shape obeying the corresponding local symmetry, manifests as an extremum in the $\kappa $ vs $\ell$ curve.

The implications of Eq.~(\ref{k_zero}) on bounded systems provides also certain interesting links to their corresponding scattering counterparts. The transition from the bounded system to the scattering one is achieved by removing the hard wall boundaries, leaving otherwise the system unaffected. Then, on either side of the setup, the potential vanishes and the wavefunction can be described by plane waves. Here, the asymptotic conditions are described by incoming and outgoing waves of the form $\psi_{I}(x)=e^{i k x}+r e^{-i k x}$ and $\psi_{III}(x)=t e^{i k x}$, respectively. Note that $k$ is the continuous wavenumber of the scattering system, while $r$ and $t$ the transmission amplitudes.

Then, we can distinguish two cases of particular interest which render Eq.~(\ref{proof5}) [and consequently Eq.~(\ref{k_zero})] also equal to zero.These cases provide the connection between special spectral points of the bounded system and the wavenumbers where PTRs occur in the transmittance of the respective scattering system. For reasons of clarity, we denote with $k_{s}$ and $k_{a}$ the wavenumbers where the $s$- and $a$-PTRs occur, respectively. Therefore, we have,

\begin{enumerate}
\item \textit{$\kappa=k_{s}$}: At this $\kappa$ value the corresponding scattering system exhibits a $s$-PTR. In the case of a $s$-PTR both potential parts are independently transparent (for a detailed analysis see Ref.~\cite{Kalozoumis2013b}). The independency refers to the fact that their distance $L$ is irrelevant to their transparency.  In terms of the TM formalism this occurs when the anti-diagonal terms are $z_{1c}(k_s)=z_{2c}(k_s)=0$. In this case ($\kappa =k_{s}$)  Eqs.~(\ref{proof5}), (\ref{k_zero}) become zero independently of the $\ell$ value. Therefore, the eigenstate with $\kappa =k_{s}$ will be invariant under translations of the setup inside the cavity and the $\kappa$ vs $ \ell/L$  curve will appear as a horizontal line. We remind here that for the scattering system the  $s$-PTR at $k_s$ will appear for any distance $L$ between the two scatterers. On the other hand, for the bounded system not all $L$ values will yield an eigenstate at $\kappa=k_s$. If however, an eigenstate with  $\kappa=k_s$ exists in the spectrum, then it will have the aforementioned translation invariance property.

\item \textit{$\kappa =k_{a}$}: At this $\kappa $ value the corresponding scattering system exhibits an $a$-PTR. In the case of an $a$-PTR the complete setup i.e. the combination of the two subparts  is transparent.  Then,  the off-diagonal terms of the total TM -which is the product $M_{\mathcal{D}_{1}} \cdot M_{\mathcal{D}_{2}}$ of the two individual TMs- is given by $z_{tot}=w_{1c}z_{2c}+z_{1c}w^{*}_{2c}e^{-i\kappa D}$, which (along with its complex conjugate)  are the quantities in the parentheses in Eq.~(\ref{proof5}).  Apparently, for the reflectionless state we have $z_{tot}=0$ and consequently Eq.~(\ref{proof5}) (and Eq.~(\ref{k_zero}) ) becomes zero.Therefore, also this $\kappa =k_{a}$ value will be unaffected by the $\ell$ variation and will appear as a horizontal line in the $\kappa $ vs $\ell$ diagram. Here, contrary to the $s$-PTR, the distance $L$ plays a major role on the transparency, since an $a$-PTR corresponds to a specific distance $L$. This leads to fact that there is always a correspondence between an $a$-PTR  of a scattering system at $k_s$ and a translation invariant bound state of its bounded counterpart at $\kappa=k_s$.  
\end{enumerate}

Inversely, it holds that a translation invariant bound state at $\kappa $ value will manifest in the transmittance of the corresponding scattering system as a PTR.

\section{Conclusions}\label{VI}

We explored a generic bounded system with hard wall boundary conditions, consisting of two locally reflection symmetric potential barriers. Focusing on the variation of the energy eigenvalues by tuning the position of the potential units inside the box, we proved the existence of spectral extrema where the mirror symmetry of the wavefunction is restored inside each locally symmetric potential barrier. These extrema accumulate to eigenenergy values which coincide with the energies where perfect transmission resonances emerge in the transmittance of the associated  scattering system. This behaviour is a benchmark of the duality between scattering and bounded systems in the presence of locally symmetric potential landscapes.
It is exemplified in this work for a system with two domains of local symmetry comprised of Dirac $\delta$ barriers. Our work could facilitate the design of cavities with prescribed spectral and wavefunction properties.
The established duality opens the perspective of linking and controlling bounded versus scattering setups. A bounded system can be designed to possess not only LS symmetric wavefunctions but its opening up to a scattering device leads also to an ``infinite range'' extension via PTRs to the outside region.

\section{acknowledgements}

The authors thank V. Achilleos for fruitful discussions. This work has been supported by the APAMAS project which is supported by "Le Mans Acoustique" and funded by the "Pays de la Loire" region and the "European Regional Development Fund".

\end{document}